# CPMD simulation of $Cu^{2+}$- phenylalanine complex under micro-solvated environment

*Aravindhan Ganesan*[a*], *Jens Dreyer*[b], *Feng Wang*[a*], *Jaakko Akola*[c,d] *and Julen Larrucea*[c]

[a]eChemistry Laboratory, Faculty of Life and Social Sciences, Swinburne University of Technology, Victoria, Australia.

[b]Computational Biophysics Laboratory, German Research School of Simulation Sciences, Jülich, Germany.

[c]Nanoscience Center, Department of Physics, University of Jyväskylä, Finland.

[d]Department of Physics, Tampere University of Technology, Tampere, Finland.

***Corresponding Author**

Telephone: +61 3 9214 5065.

Fax: +61 3 9214 5921.

Email: aganesan@daad-alumni.de

**Abstract**

The study combines density functional theory calculations and Car-Parrinello molecular dynamics (CPMD) simulations to investigate the structures of phenylalanine-copper (II) ($[Phe\text{-}Cu]^{2+}$) complexes and the micro-solvation processes. The structures of the $[Phe\text{-}Cu]^{2+}$ complex with up to four water molecules are optimized quantum mechanically using the B3LYP/6-311++G** model in gas phase to identify the lowest energy structures at each level of solvation (n=0-4). It is found that the phenylalanine moiety appears to be in the neutral form in isolated and mono-hydrated complexes, but in the zwitterionic form in other hydrated complexes (with $n \geq 2$). The energy minimum structures of the complexes suggest that the $Cu^{2+}$–$\pi$ interactions are not dominant in the $[Phe\text{-}Cu]^{2+}$ complexes. The present CPMD simulations of the lowest energy micro-hydrated $[Phe\text{-}Cu]^{2+}$ complexes also reveal that the maximum coordination of $Cu^{2+}$ in the presence of the Phe ligand does not exceed four: the oxygen atoms from three water molecules and one carboxyl oxygen atom of Phe. Any excess water molecules will migrate to the second solvation shell. Moreover a unique structural motif, $(N)H\cdots O_{(3)}\cdots H_2O\text{–}Cu^{2+}$ is present in the lowest energy complexes, which is recognized to be significant in stabilizing the structures of the complexes. Extensively rich information of the structures, energetics, hydrogen bonds and dynamics of the lowest energy complexes are discussed.

## Introduction

Amino acids, metal ions and waters are among the most important chemical partners in the biology of life as the interactions between these components play a crucial role in the structures and functions of proteins in aqueous solution [1]. Metal interactions with amino acids are significant in electron transfer or deprotonation reactions [2-4] within the molecular systems, and also affect protein folding/unfolding and aggregation processes [5]. Studying the interactions of solvent molecules with amino acids and metal ions is thus vital for understanding the hydration of metalloproteins and the role of water molecules in biological systems [1]. Although bulk solvents are generally considered when studying bio-molecular systems, micro-solvation effects can potentially change molecular properties [6, 7].

Micro-solvation effects play significant roles in diverse molecular systems, ranging from amino acids to large enzymes. The neutral (NT)-to-zwitterion (ZW) switching of amino acids takes place in the presence of micro-solvents [8]. Rodziewicz and Doltsinis [8] showed in their *ab initio* molecular dynamics (AIMD) investigation on micro-hydrated phenylalanine (Phe) structures that more than three water molecules are required to fully stabilize the ZW forms of Phe in water. However, the ZW form of Phe, when complexed with the aluminum metal ion, was found to be stabilized with only two water molecules [9]. Therefore the number of water molecules required for conformational transitions of amino acids remains an open question. Moreover a recent work by Otto et al [10] published in Nature Chemistry, unveils the influence of micro-solvents in the reaction dynamics of hydroxyl ions with iodomethane using a combined crossed-beam imaging and cold source of micro-solvated reactant experiment and ab-initio simulations. The study finds that distinct reactions take place in different degrees of solvation, while the co-linear substitutions happen under mono-solvation [10]. Micro-hydration studies can therefore provide subtle information on the solute-solvent interactions that can be useful for a molecular level understanding of the complex systems.

The significance of transition metal-ligand complexes has been recognized for decades in areas such as catalysis, drug research, atmospheric chemistry and biochemistry [11]. A number of previous experimental and theoretical studies on the interactions between several transition metal ions have been reported [2, 11-23]. Copper is one of the most prominent transition elements in biological systems [16, 24], with an occurrence of 80-120 mg in a normal human body [16]. Copper is also a component of several enzymes such as indophenoloxidases, cytochrome *c* oxidase, Cu/Zn superoxide dismutase and tyrosinase [25]. In the current work, we study the effects of copper (II), a transition metal ion, on the structures of Phe under the micro-solvated environment.

The $Cu^{2+}$ ion with nine *d* electrons ($d^9$) in its valence shell in the electronic configuration of $(t_g)^6$ $(e_g)^3$, often displays an octahedral coordination in crystals and aqueous solution. The four equatorial and two axial bonds have been well described by a number of experiments [16, 26-35] including X-ray absorption [29], NMR [32], X-ray absorption near-edge structure (XANES) [28, 30, 31], neutron diffraction [16, 26, 35] and extended X-ray absorption fine structure (EXAFS) [27, 28, 30, 33, 34]. The two axial bonds in the $Cu^{2+}$ within aqueous solution are generally elongated due to the Jahn-Teller distortion effects [33, 34]. In contrast, Pasquarello et al. [16] have deduced from their combined neutron diffraction and AIMD simulations that $Cu^{2+}$ ions favor fivefold trigonal bipyramidal configurations in aqueous solution. This was later supported by an X-ray absorption spectroscopy based analysis [28]. However, the solid $Cu^{2+}$ complexes with aliphatic amines, pyridines [36] or ammonia [37-39] have been recognized to show a variety of different coordination: square planar (fourfold) [40, 41], square pyramidal (fivefold)[42, 43] and distorted square bipyramidal (sixfold) [44, 45]. Rulisek and Vondrasek [46], who have exploited several metalloproteins and transition metal complexes from the Protein Data Bank (PDB) and the Cambridge Structure Database (CSD), ascertained that the $Cu^{2+}$ metal ion mostly prefer a square planar structure, although a few complexes also exhibit trigonal bipyramidal geometries. Hence the coordination of $Cu^{2+}$ within different molecular environments is often debatable. Previous studies discussed the coordination of $Cu^{2+}$ metal ions with aromatic amino acids. For example, Rimola et al. [22, 24] used density functional theory (DFT) based theoretical calculations to study the interactions of $Cu^{+}$ and aromatic amino acids. Remko et al. [17] studied $Cu^{2+}$ – aromatic amino acid complexes in the gas phase and in the presence of 5 water molecules using DFT. It was found that the copper-aromatic amino acid system exists as an NT conformer and the copper-aromatic amino acid-5 water system exists as a ZW conformer.

The effects of stepwise micro-hydration of $Cu^{2+}$–Phe complexes (i.e., $[Phe\text{-}Cu]^{2+}$) are limited in the literature. Micro-hydration studies of $[Phe\text{-}Cu]^{2+}$ can reveal important information that are usually hidden in the fully solvated systems. The present work uses DFT calculations combined with Car-Parrinello molecular dynamics (CPMD) simulations to investigate the effects of $Cu^{2+}$ binding on the Phe structures under micro-solvated environments, $(H_2O)_{n=1-4}$. The geometry optimizations are performed to identify the lowest energy structures for all $[Phe\text{-}Cu(H_2O)_{n=1-4}]^{2+}$ complexes, followed by CPMD simulations to probe structural changes at room temperature and to reveal the coordination preferences of $Cu^{2+}$ ion for different micro-hydrated states.

## Methods and computational details

The $Cu^{2+}$–Phe complexes are built using the configurations from Larrucea et al. [9] on $Al^{2+}$–Phe complexes as reference structures. In general, there are a number of approaches such as simulated annealing, exhaustive stochastic search, etc., for identifying stable conformers of biomolecules and molecular clusters [47, 48]. Nevertheless, the different conformations of amino acids, which are stable in the gas phase, have been reported earlier [49-51]. For the present work, we chose five different structures of L-Phe from the literature [49-51], Phe1, Phe2 and Phe3, which are ground state NT configurations with –COOH/–$NH_2$ groups and Phe4 and Phe5, which are ZW conformers with –$COO^-$/–$NH_3^+$ groups. During the model building process, the $Cu^{2+}$ ion is placed at different binding positions around the parent structures, in order to construct different Phe-$Cu^{2+}$ complexes. Both the NT and ZW complexes contain structures with the $Cu^{2+}$ metal ion binding to carboxyl-amino groups as well as the phenyl ring. The formation of $[Phe\text{-}Cu]^{2+}$ complexes (i.e., n=0) can be described by the following process,

$$Phe_{(NT/ZW)} + Cu^{2+} \longrightarrow [Phe_{(NT/ZW)}\text{-}Cu]^{2+} \qquad (a)$$

Next, water molecules are added one by one to the n=0 complex around the $Cu^{2+}$ ion to form the micro-hydrated systems that include n=1, 2, 3 and 4 complexes according to the hydration reaction,

$$[Phe_{(NT/ZW)}\text{-}Cu(H_2O)_i]^{2+} + H_2O \longrightarrow [Phe_{(NT/ZW)}\text{-}Cu(H_2O)_{i+1}]^{2+} \qquad (b)$$

where i=0, 1, 2 and 3. As a result, a total of 35 structures as initial structures of the complexes are produced. A number of tools, such as Molden [52] and VMD [53] are used to model and visualize the structures in this study.

Geometry optimizations of all 35 initial structures are performed using two different DFT functionals, B3LYP and BLYP. The calculations with B3LYP together with the 6-311++G(d,p) basis set are carried out using the latest version of the Gaussian09 (G09) [54] computational chemistry package. For the open shell $Cu^{2+}$, we employ the Watcher-Hay all electron basis set [55, 56] that is able to accurately describe the energies of the transition metal ions [17]. The hybrid B3LYP model provides accurate geometries as found in our previous studies involving a number of amino acids [57-59] and has also been proven to be efficient for many transition metal-containing systems [11, 60-62].

The CPMD [63, 64] simulations are performed for the lowest energy micro-hydrated systems (i.e., $_{l}$(n=1-4)). A periodic cubic box of 18Å in length is employed. The default Hockney Poisson solvers and the local density approximation (LDA) are employed, whereas, the valence electrons are treated explicitly using the BLYP functional. The exchange functional is given by Becke [65] and the correlation energy expression by Lee, Yang and Parr [66] in the BLYP functional. The core electrons are described using the norm-conserving Troullier and Martins [67] pseudopotentials. For $Cu^{2+}$, the non-linear core correction (NLCC) [68] pseudopotential is used to improve the description of its core energy. This NLCC correction has been recognized to be more appropriate for the transition metal ions [21]. The Kohn-Sham orbitals are expanded with a plane wave cut-off of 90 Ry. An electronic fictitious mass of 600 amu with a time step of 5.0 a.u. is used. Under these conditions, the molecular dynamics simulations of the $_{l}$(n=1-4) systems are performed for more than 12 ps. The temperature is maintained at 300 K throughout the simulation using a Nosé-Hoover [69, 70] thermostat. Further we imply a geometric criteria of 2.80 Å cut-off for determining the inter- and intra-molecular H-bonds in our systems [71]. All the parameters described above are chosen following a series of convergence tests carried out using the CPMD package.

## Results and discussions

### Optimized geometrical structures

Hydration effects on bare $Cu^{2+}$ metal ions are studied at first to validate the theoretical models used in this work. The optimized structures along with their energies are given in supplementary information (S1). The bond distances in these structures agree very well with those obtained in previous studies [20, 72] as well as with experimental data [16, 27, 34] (Table S2). Specifically, the averaged Cu-O and O-O distances in the first and second hydration shells are in excellent agreement with the experimental values. For example, the averaged Cu-O distances measured by various experiments are between 1.94-2.15 Å, in fairly good agreement with the present calculations of 1.85-2.11 Å. Similarly, the O-O distances in our calculations are 2.69 Å and 2.71 Å, which are in reasonable agreement to the experimental values (2.73-2.80 Å) [73, 74]. As a result, both theoretical models, B3LYP/6-311++G(d,p) and BLYP/MT, are able to satisfactorily represent $Cu^{2+}$ complexes.

Optimized geometries of the $[Phe\text{-}Cu(n\text{=}1\text{-}4)]^{2+}$ structures considered here are shown in Fig. 1. The first row gives the conformers of Phe in isolation, the next row presents the $[Phe\text{-}Cu]^{2+}$ complexes (i.e. n=0) produced from the Phe conformers in the first row with $Cu^{2+}$, but without water molecules; whereas rows 3-6 display the hydrated $[Phe\text{-}Cu]^{2+}$ complexes with waters (n=1-4). Harmonic vibrational frequency analyses indicate that these complexes are true minimum structures. Energies of the complexes are corrected for basis set superposition errors (BSSE), which are calculated using the counterpoise methods implemented in G09. The relative energies calculated using the B3LYP/6-311++G(d,p) model are given with the complexes in Fig. 1.

There are five Phe conformers (Fig. 1, first row) in isolation, which comprise three conformers in NT form (Phe1-Phe3) as well as two ZW conformers (Phe4-Phe5). Conformer Phe3 is the lowest energy structure among them, in agreement with previous studies.[58, 71] As expected, the relative energies of the gas phase ZW structures, Phe4 and Phe5, are much higher in energy, 16.5 $kcal\cdot mol^{-1}$ and 15.7 $kcal\cdot mol^{-1}$ respectively, than the NT Phe3 minimum structure. The second row in Fig. 1 displays the $[Phe\text{-}Cu]^{2+}$ complexes where two different metallated complexes are obtained for the ZW structure Phe5, i.e., $[Phe5\text{-}Cu]^{2+}$ and $[Phe5a\text{-}Cu]^{2+}$. In the former ($[Phe5\text{-}Cu]^{2+}$), the $Cu^{2+}$ binds with the carboxyl moiety but in the latter the $Cu^{2+}$ ion also binds other carbon atoms in the phenyl ring.

Table 1 compares the selected geometrical parameters of all the $[Phe\text{-}Cu]^{2+}$ complexes calculated using B3LYP (G09) and BLYP (CPMD) methods along with results from other studies.[17] The geometries of $[Phe1\text{-}Cu]^{2+}$ and $[Phe4\text{-}Cu]^{2+}$ show excellent agreement with those reported recently by Remko et al[17] (calculated using B3LYP/6-311+G** model), although the models using slightly different basis sets. Comparison of the different $[Phe\text{-}Cu]^{2+}$ complexes in this study show very different geometrical parameters. This is particulary the case when $Cu^{2+}$ is involved. For example, $O_{(4)}$-Cu and $O_{(3)}$-Cu bond lengths, $C_{(2)}$-$O_{(4)}$-Cu and $C_{(2)}$-$O_{(3)}$-Cu bond angles and $C_{(1)}$-$C_{(2)}$-$O_{(4)}$-Cu and $C_{(1)}$-$C_{(2)}$-$O_{(3)}$-Cu dihedral angles, display larger deviations among the complexes shown in Table 1. See footer in fig. 1 for the atom numbering in the NT and ZW forms of Phe. Moreover, the ring perimeters ($R_6$) of the phenyl rings within the Phe structures listed in the Table 1 are longer than the $R_6$ of the isolated global minimum Phe of 8.37 Å.[58] As a result, upon $Cu^{2+}$ binding the phenyl ring expands so that the $R_6$ of the complexes are longer.

Of all the $[Phe\text{-}Cu]^{2+}$ complexes considered here (Fig.1) the most stable structure is the $[Phe1\text{-}Cu]^{2+}$ complex, in which the metal ion forms a bidentate coordination with the amino nitrogen (N) and the carboxyl oxygen ($O_{(3)}$) atoms. It is realized that the electrostatic interactions and repulsions from the strongly electronegative oxygen (especially carbonyl oxygen) and nitrogen atoms in Phe1 is more favorable for the $Cu^{2+}$ ion. This agrees with previous studies[11, 17, 22-24, 75] showing that the transition metal ions mostly bind with the N and O atoms of aliphatic and aromatic molecules in order to present stable complex structures in the gas phase. The ZW complexes, $[Phe4\text{-}Cu]^{2+}$ and $[Phe5\text{-}Cu]^{2+}$ are found to be the next preferred complexes with only 4.6 kcal·mol$^{-1}$ and 4.5 kcal·mol$^{-1}$ higher in energy, respectively, than the most stable NT complex, $_{l}[Phe1\text{-}Cu]^{2+}$.

Addition of the first water molecule (n=1) to the $[Phe\text{-}Cu]^{2+}$ complex does not lead to any considerable structural changes, as was also observed in other metal-aromatic amino acid systems.[9, 76] The same NT conformer is still preferred as the lowest energy complex upon mono-hydration (i.e., $_{l}$( $[Phe1\text{-}Cu(n=1)]^{2+}$ in the third row of Fig.1), where the water molecule is attached to the NO-coordinated $Cu^{2+}$ ion and without any direct contact to Phe.

Introducing the second water molecule into the complex leads to the ZW structures being energetically preferred with respect to the NT structures. In the lowest energy (n=2) complex (i.e., $_{l}[Phe4\text{-}Cu(n=2)]^{2+}$ complex, the $Cu^{2+}$ metal ion interacts with both the carboxyl oxygen atoms ($O_{(3)}$ and $O_{(4)}$) of the Phe moiety and the two water molecules (Fig.1). One of the other ZW complexes, $[Phe5a\text{-}Cu(n=2)]^{2+}$, is only 1.6 kcal·mol$^{-1}$ higher in energy than the $_{l}[Phe4\text{-}Cu(n=2)]^{2+}$ complex. The NT complexes are now much higher in energy than the ZW ones, with $[Phe1\text{-}Cu(n=2)]^{2+}$ being 5.1 kcal·mol$^{-1}$ higher than the most stable one, while the other NT structures are approximately 10.9 kcal·mol$^{-1}$ and 15.3 kcal·mol$^{-1}$ higher in energy. This indicates that two water molecules are sufficient to interconvert the energetic order of the NT and ZW configurations of $[Phe\text{-}Cu]^{2+}$ complexes in the micro-hydration processes. Such a conversion from NT to ZW was experimentally confirmed[77] in $[Val(Na)]^+$ in the presence of two water molecules, but no such experimental evidence is available for the $[Phe\text{-}Cu]^{2+}$ systems. Earlier CPMD based investigation[8] reveals that the Phe structure, without any metal ions, require larger numbers of water molecules ($n \geq 3$), for stabilizing the ZW form of Phe, whereas the present study finds that ($n=2$) is sufficient to deliver the NT → ZW transformation. This indicates that the presence of $Cu^{2+}$ may have changed the mechanism and reduced the complexity in the de-protonation process of Phe.

Upon addition of the third water molecule, the ZW complex $[Phe5\text{-}Cu(n=3)]^{2+}$ is identified as the most stable structure, i.e. $_{l}[Phe5\text{-}Cu(n=3)]^{2+}$. In this complex, the $Cu^{2+}$ ion shows a distorted square planar coordination including a carboxyl oxygen, $O_{(4)}$, and the oxygen atoms of three water molecules. The energy for the second most stable structure, $[Phe4\text{-}Cu(n=3)]^{2+}$, is only 1.9 kcal·mol$^{-1}$ above the $_{l}[Phe5\text{-}Cu(n=3)]^{2+}$ structure. In the $_{l}[Phe5\text{-}Cu(n=3)]^{2+}$ structure, $Cu^{2+}$ displays a penta-coordination, i.e. two with the carboxyl oxygens and the other three with water molecules. Similarly, in the most stable four water system ($_{l}[Phe5a\text{-}Cu(n=4)]^{2+}$), $Cu^{2+}$ interacts with $O_{(4)}$ of Phe and all four water molecules. The $[Phe4\text{-}Cu(n=4)]^{2+}$ complex again with a penta-coordinated $Cu^{2+}$ metal ion is the second most stable structure and is only 0.4 kcal·mol$^{-1}$ higher in energy than $_{l}[Phe5a\text{-}Cu(n=4)]^{2+}$.

A few complexes were initially built with $Cu^{2+}$-phenyl bonds in order to study the cation-π interactions. However during geometry optimization, the cation-π interactions in the complexes broke and evolved into structures, in which the $Cu^{2+}$ moves away from the phenyl ring, thus avoiding the π interactions. Despite a few optimized complexes such as $[Phe5\text{-}Cu]^{2+}$, $[Phe5a\text{-}Cu(n=1)]^{2+}$ and $[Phe3\text{-}Cu(n=2)]^{2+}$ for instance, still hold the cation-π bonds, their relative energies are high. The $Cu^{2+}$-phenyl bonds do not appear in the most stable structures. This is in contrast with a previous ab-initio study[8] in which, the water···π hydrogen bonds (H-bonds) are crucial for stabilizing the micro-hydrated Phe systems (without a $Cu^{2+}$).[8] It is noted that the hydrated complexes formed with the global minimum structure of Phe (i.e., Phe3) become the least stable complexes. This suggests that the inter-molecular forces of hydrated complexes can change the complexes considerably from gas phase.

Table 2 presents representative $Cu^{2+}\text{-}O_{(Phe)}$ and $Cu^{2+}\text{-}O_{(wat)}$ distances for the lowest energy complexes ($_{l}(n=0\text{-}4)$). The $Cu^{2+}$-O distances in $_{l}[Phe1\text{-}Cu]^{2+}$ and $_{l}[Phe1\text{-}Cu(n=1)]^{2+}$ can be compared directly, as both complexes are based on Phe1, with $Cu^{2+}$ ion interacting with the nitrogen atom and the carboxyl oxygen atom, $O_{(4)}$, of the Phe moiety (i.e., N-$Cu^{2+}$-$O_{(4)}$). In the absence of water, i.e., in the $_{l}$[Phe1-Cu]$^{2+}$ complex, the $Cu^{2+}$ cation binds tightly with $O_{(4)}$ of Phe. When one water molecule is introduced, the binding between $Cu^{2+}$ and the Phe loosens in the $_{l}$[Phe1-Cu(n=1)]$^{2+}$ complex. For example, the distance between $Cu^{2+}$ and $O_{(4)}$ in the $_{l}$[Phe1-Cu]$^{2+}$ complex is given by 2.05 Å using the B3LYP/6-311++G(d,p) model, which increases to 2.09 Å in the $_{l}$[Phe1-Cu(n=1)]$^{2+}$ complex upon addition of a water molecule. The increase of the $Cu^{2+}$-$O_{(4)}$ (Phe) distance comes at the expense of the Cu-N binding. For example, the Cu-N distance in the $_{l}$[Phe1-Cu]$^{2+}$ complex is given by 2.08 Å, but reduced to 2.01 Å in the $_{l}$[Phe1-Cu(n=1)]$^{2+}$ complex. Such shortening of the Cu-N distance has also been reported by Remko et al. [17] Nevertheless, the $Cu^{2+}$-$O_{(3)}$ distance is always larger than the $Cu^{2+}$-$O_{(4)}$ distance. The calculated water-$Cu^{2+}$ distances in most of the lowest energy complexes are within the experimentally observed range of 1.94 – 2.15 Å [20, 21, 34, 72] for bare $Cu^{2+}(H_2O)_n$ (without Phe).

**CPMD simulations**

The CPMD simulations of the lowest energy micro-hydrated structures, $_{l}$[Phe1-Cu(n=1)]$^{2+}$, $_{l}$[Phe4-Cu(n=2)]$^{2+}$, $_{l}$[Phe5-Cu(n=3)]$^{2+}$ and $_{l}$[Phe5a-Cu(n=4)]$^{2+}$ are performed at room temperature for ~12 ps. The snap-shots of the structures taken at the end of ~12 ps of CPMD simulation are given. Significant geometrical parameters of the initial and final configurations (i.e., the configuration of the last snapshot) of the complexes are compared in Table 3. The $_{l}$[Phe1-Cu(n=1)]$^{2+}$ complex shown in Fig. 2a remains quite stable in its NT configuration during the MD simulation, with only few changes in its bond lengths. For instance, in the initial structure of $_{l}$[Phe1-Cu(n=1)]$^{2+}$, the Cu-N bond (2.03 Å) is shorter than the Cu-$O_{(4)}$ bond (2.12 Å); whereas in the final structure, the situation is inversed. That is, the Cu-N bond becomes larger than the Cu-$O_{(4)}$ bond in the final structure, in agreement with an earlier study [17]. It is found that the coordination of Cu-$O_{(wat1)}$ is weakened slightly as the bond distance increases from 1.98 Å to 2.09 Å. This indicates that the inter-conversion between the NT and ZW transition of the $_{l}$[Phe1-Cu(n=1)]$^{2+}$ complex is unlikely to occur spontaneously.

The $Cu^{2+}$ ion in the initial structure of the $_{l}$[Phe4-Cu(n=2)]$^{2+}$ complex displays a square-planar coordination, a pair to the carboxyl oxygen and the other pair to the water molecules. However, during the dynamics process, one of the Cu-O bonds such as the Cu-$O_{(3)}$ bond weakens and the $Cu^{2+}$ ion switches to a tridentate coordination (fig. 6.3(b)). This is reflected in the changes in bond lengths (table. 3), that is, the Cu-$O_{(3)}$ distance of the initial structure is given by 2.18 Å; whereas in the final structure, this distance increases to 3.19 Å. Consequently, the $O_{(3)}$ atom is replaced by a H-bond with a water molecule, in addition to its pre-existing intra-molecular H-bond with the amino group (i.e., (N)H···$O_{(3)}$). Thus both intra- and inter-molecular interactions together form a (N)H···$O_{(3)}$···$H_2O$–$Cu^{2+}$ chemical bond network within the $_l$(n=2) complex. Here, a water molecule serves as a bridge between the carbonyl oxygen ($O_{(3)}$) atom and the $Cu^{2+}$ atom ($O_{(3)}$···$H_2O$–$Cu^{2+}$). Similar H-bonded bridges between the $Cu^{+}$ cation and the curcumin anion have been reported very recently by Addicoat et al [78]. Breaking of the Cu-$O_{(3)}$ bond leads to a stronger coordination between the $Cu^{2+}$ and $O_{(4)}$ atom reflected by the reduction of its distance from 2.08 Å to 1.94 Å.

In the $_l$[Phe5-Cu(n=3)]$^{2+}$ complex, the structural rearrangement is similar to that of the bi-hydrated complex (Fig. 2c). However, binding of an additional water molecule (wat3) to $Cu^{2+}$ changes the coordination of $Cu^{2+}$ to square-planar. This additional water molecule is H-bonded to the $O_{(3)}$ atom of Phe, thereby preventing it from directly binding to the $Cu^{2+}$ ion. Hence the $_l$[Phe5-Cu(n=3)]$^{2+}$ complex also adapts the (N)H···$O_{(3)}$···$H_2O$–$Cu^{2+}$ chemical bond network after the interactions with three water molecules, which is found by the molecular dynamics simulation.

The restructuring of the $_l$[Phe5a-Cu(n=4)]$^{2+}$ complex is revealed by the CPMD simulation, as shown in Fig. 2d. The structure of the $_l$[Phe5a-Cu(n=4)]$^{2+}$ complex is more complicated than other hydrated systems with less water molecules involved (i.e. $n<4$), as discussed above. The $Cu^{2+}$ ion in the system starts to exhibit a distorted penta-coordinated trigonal-bipyramidal configuration, where the $Cu^{2+}$ atom is coordinated with all the four water molecules and one carboxyl oxygen, $O_{(4)}$. The water molecules around $Cu^{2+}$ can cause increased steric hindrance. A very recent study by Otto et al [10] again indicates the important role of the steric characteristics of the water molecules in micro-solvated reactions. This can be the case at room temperature as revealed by the CPMD simulation. The rearrangement of the complex to create a favorable steric environment in the system is found. As a result, one of the water molecules (wat4), which is initially connected to the $Cu^{2+}$ cation, is moved to the second coordination shell thereby reducing the 'steric attack' towards $Cu^{2+}$ and resulting in a square-planar like complex.

Water migrations between different solvation shells may affect the inter- and intra-molecular interactions of the complex. As shown in Table 3, the initial structure of the $_{l}$[Phe5a-Cu(n=4)]$^{2+}$ complex possesses only two types of H-bonds, (N)H···$O_{(3)}$ and $H_{(wat3)}$···$O_{(3)}$. Additional H-bonds between the water molecules, wat1, wat2 and wat4, are formed in the final complex after the dynamical process. Note that in the initial $_{l}$[Phe5a-Cu(n=4)]$^{2+}$ complex, the (N)H···$O_{(3)}$ H-bond exists between the $Cu^{2+}$ ion and the $H_x$ atom in the $NH_3$ group of ZW Phe, while in the final structure, the $H_y$ atom in the $NH_3$ forms H-bond with $Cu^{2+}$. Such a change in (N)H···$O_{(3)}$ bond is due to the rotation of the $NH_3$ group during the CPMD simulations. Here, Hx and Hy indicate two different hydrogen atoms in the amino group ($NH_3$) of the phenylalanine moiety. More H-bonds in the final configuration indicate that the $_{l}$[Phe5a-Cu(n=4)]$^{2+}$ complex is stabilized by formation of the H-bonding network among water molecules in different solvation shells, Phe and $Cu^{2+}$. Moreover, similar to the bi- and tri-hydrated systems, one of the water molecules remains as a coordination mediator between the $O_{(3)}$ atom and the $Cu^{2+}$ ion (see Fig. 2d). The (N)H···$O_{(3)}$···$H_2O$–$Cu^{2+}$ H-bond chain, therefore, remains a unique structural motif to stabilize the lowest energy [Phe-Cu]$^{2+}$ complexes with more than two water molecules.

Fig. 3 presents dynamical trajectories of the Cu-O distances (Fig. 3a) and the H-bonds (Fig. 3b and c) of the $_{l}$[Phe5a-Cu(n=4)]$^{2+}$ complex for a period up to 12ps. Fig. 3a shows that the Cu-O distances of Cu-$O_{(wat1)}$, Cu-$O_{(wat2)}$, Cu-$O_{(wat3)}$ and Cu-$O_{(4)}$ are stable with the variations between 2-2.5 Å. As shown in the structure of the complex given in Fig. 3, the $Cu^{2+}$ directly bonds with oxygen atoms presented in the four bonds. As a result, these Cu-O bonds are strong and dynamically stable in this period of time. Another dynamically stable but weaker Cu···O distance shown in Fig. 3a is the Cu···$O_{(3)}$ "bond", which remains in the vicinity of 3.5 Å. This Cu···$O_{(3)}$ "bond" is stabilized by the formation of a H-bond with a hydrogen atom of the water molecule (wat3) that is directly bonded with the $Cu^{2+}$ ion (i.e., Cu-$O_{(wat3)}$). This water molecule (wat3) "bridges" the $O_{(3)}$ atom of Phe with $Cu^{2+}$ and makes the squared-planar $Cu^{2+}$ configuration of the complex possible without causing a significant strain to the complex, as $O_{(4)}$ of Phe already bonds with the metal.

Fig. 3a displays the significant dynamical changes in the $_{l}$[Phe5a-Cu(n=4)]$^{2+}$ complex The Cu··· $O_{(wat4)}$ distance. The Cu···$O_{(wat4)}$ bond is the only dynamically unstable Cu···O bond of the $_{l}$[Phe5a-Cu(n=4)]$^{2+}$ complex. This Cu···$O_{(wat4)}$ bond undergoes significant changes from being a strong Cu-O bond with the distance of < 2.5 Å to a weak Cu···O bond with the distance of ~4.5 Å. The dynamical change over happens at ~6 ps, as shown in Fig. 3a. The oxygen atom of this water molecule (wat4) is initially bound with $Cu^{2+}$, forming a strong bond of Cu-$O_{(wat4)}$. However, the metal ion prefers a squared-planar configuration after 6 ps, rather than the initial distorted penta-coordinated trigonal-bipyramidal configuration. As a result, the Cu···$O_{(wat4)}$ distance starts to increase and stabilizes at ~4.5 Å. The $O_{(wat4)}\cdots O_{(wat2)}$ distance in the final structure is given by 2.60 Å, which is in good agreement with the experimental value of 2.73 Å for the oxygen-oxygen distance between the two water molecules in the first and second solvation shell, respectively [34]. The significant increase in the Cu···$O_{(wat4)}$ distance, therefore, indicates the substantial changes in the configuration of the metal complex, which releases the fourth water molecule (i.e. wat4) to the second solvation shell of the $Cu^{2+}$ ion.

Fig. 3b and 3c show that the changes in the Cu-O bonds also affect the inter- and intra-molecular H-bonds of Cu···$O_{(wat4)}$ bond. It can be seen that the inter-molecular H-bond, $H_{(wat3)}\cdots O_{(3)}$ remains stable at < 2.5 Å throughout the simulation thereby bridging between Phe and $Cu^{2+}$ (Fig. 3b). The intra-molecular H-bond between the amino group and the carboxyl group (i.e. (N)H···$O_{(3)}$) of Phe (Fig. 3c) exists throughout the simulation. However, a switch over between the (N)$H_x\cdots O_{(3)}$ in the initial structure to (N)$H_y\cdots O_{(3)}$ in the final structure. Here again the $H_x$ and $H_y$ represent two different hydrogen atoms in the amino group of the Phe ZW moiety. Fig. 3c clearly indicates that the $H_x\cdots O_{(3)}$ and $H_y\cdots O_{(3)}$ change over occur in simultaneous with the changes of the Cu···$O_{(wat4)}$ distances, as shown in Fig. 3a. The result of such structural changes in the $_l$[Phe5a-Cu(n=4)]$^{2+}$ complex leads to the (N)H···$O_{(3)}\cdots H_2O$–$Cu^{2+}$ H-bond chain, which is similar to the other hydrated complexes, $_l$[Phe4-Cu(n=2)]$^{2+}$ complex and $_l$[Phe5-Cu(n=3)]$^{2+}$ complex, discussed previously in the text. The other inter-molecular H-bond, $H_{(wat2)}\cdots O_{(wat4)}$, starts to form at almost the same time of the Cu···$O_{(3)}$ and NH···$O_{(3)}$ changes in the complex, i.e., at approximately 6 ps, as shown in Fig. 3b. That is, at ~6ps when the penta-coordinated trigonal-bipyramidal configuration of the $Cu^{2+}$ ion (which is the configuration of the bare $Cu^{2+}$ metal ion under full solvation [16, 21]), starts to take the squared planar configuration in the $_l$[Phe5a-Cu(n=4)]$^{2+}$ complex with phenylalanine. As such, the Cu-$O_{(wat4)}$ bond becomes a weaker Cu···$O_{(wat4)}$ interaction, which allows the wat4 molecule to migrate towards the second solvation shell. As a result, Cu-$O_{(wat4)}$ bond becomes a weaker Cu···$O_{(wat4)}$ interaction to allow the wat4 molecule to migrate towards the second solvation shell. In the presence of Phe, the $Cu^{2+}$ prefers to confine itself to a distorted square planar coordination in order to reduce the steric hindrance and maintain its interactions with the Phe under micro-hydration.

Fig. 4 shows the Cu-O radial distribution functions (RDFs) for the hydrated $[Phe\text{-}Cu]^{2+}$ complexes that are calculated from the CPMD trajectories. Two distinctive peaks appear in the RDF spectra of the micro-hydrated complexes. An intense peak at ~2.0 Å of the complexes is due to the strong $Cu^{2+}$-O bonds, i.e., Cu-$O_{(wat1)}$, Cu-$O_{(wat2)}$, Cu-$O_{(wat3)}$ and Cu-$O_{(4)}$, in agreement with our previous discussions. The less intense peak representing the Cu-$O_{(3)}$ distance occurs at ~3.5 Å in the complexes, except for the $_l[Phe1\text{-}Cu(n=1)]^{2+}$ complex, in which this less intensive peak locates at approximately 4.2 Å. It is this peak (i.e., the less intensive peak at 4.2 Å) which reflects the structural differences of the Phe moiety (NT or ZW) in the complexes. The Phe moiety in the $_l[Phe1\text{-}Cu(n=1)]^{2+}$ complex exhibits an NT conformer (O=C-OH), whereas the Phe moiety in other hydrated complexes exists as the ZW form ($O=C\text{-}O^-$). As a result, the RDF spectra provide useful information to differentiate the Phe NT from the ZW form of the complexes.

In addition, the RDF spectra in this work further indicates that the $O_{(3)}$ atom in the NT Phe moiety of the $_l[Phe1\text{-}Cu(n=1)]^{2+}$ complex is unlikely to directly interact with the $Cu^{2+}$ ion. The ZW Phe of the $_l[Phe4\text{-}Cu(n=2)]^{2+}$, $_l[Phe5\text{-}Cu(n=3)]^{2+}$ and $_l[Phe5a\text{-}Cu(n=4)]^{2+}$ complexes can be stabilized if the $O_{(3)}$ of Phe indirectly forms a network with $Cu^{2+}$ ion via a water bridge. Very small peaks of the $_l[Phe5a\text{-}Cu(n=4)]^{2+}$ complex at larger distances (> 4 Å) are likely due to weak interactions caused by the wat4 molecule in the second solvation shell. Again, the RDF spectra of the complexes further describes that the $[Phe\text{-}Cu]^{2+}$ complexes prefer to bond up to three water molecules directly due to steric environment of the $Cu^{2+}$-Phe network. Any excess water molecules are unable to form sufficiently strong bonds directly with the $Cu^{2+}$ ion and move to the subsequent solvation shell.

## Conclusions

Mechanism of the micro-solvation processes ($H_2O$, n=1-4) for the $Cu^{2+}$-Phe complexes has been studied using combined DFT calculations as well as CPMD simulations. In the micro-solvation processes, the number of water molecules is added one by one, starting at n=0, to form the complexes. The present study demonstrates that the complex is saturated by a maximum of four water molecules, which result in a total of thirty-five stable complexes. The lowest energy structures at each micro-hydrated level are obtained using the DFT baed geometry optimization and are further studied using the CPMD simulation at room temperature.

It is found that the number of water molecules involved in the hydrated complexes has an apparent impact on the configuration of the complexes. The present study reveals that a minimum of two water molecules are required to assist the inter-conversion of the Phe moiety between its NT and ZW configurations in the Phe-$Cu^{2+}$ complex. The complex prefers a NT Phe in the configuration of the Phe-$Cu^{2+}$ molecules without water, $_l$[Phe1-Cu]$^{2+}$, or with a single water molecule, $_l$[Phe1-Cu(n=1)]$^{2+}$. The Phe moiety prefers the ZW if more water molecules are presented, i.e. $_l$[Phe4-Cu(n=2)]$^{2+}$ , $_l$[Phe5-Cu(n=3)]$^{2+}$ and $_l$[Phe5a-Cu(n=4)]$^{2+}$.

The present study further reveals that the micro-hydration mechanisms with the presence of the $Cu^{2+}$ in the Phe-$Cu^{2+}(H_2O)_n$ complexes are very different from the Phe$(H_2O)_n$ complexes (in the absence of $Cu^{2+}$) through 'cation-π' interactions found by an earlier study [8]. The most likely contact sites for $Cu^{2+}$ includes the carboxyl oxygen atom of Phe and the oxygen atoms from up to three water molecules to form the squared planar coordination of $Cu^{2+}$ ion in the most stable $Cu^{2+}$-Phe complexes. The (N)H$\cdots O_{(3)}\cdots H_2O$–$Cu^{2+}$ network has been identified in the present study to play a significant role in the stabilization of the micro-solvated $Cu^{2+}$-Phe complexes.

## Acknowledgements

FW acknowledges her Vice-Chancellor's Research Award at Swinburne University of Technology and the ARC Centre of Excellence for Antimatter-Matter studies, Flinders University node, for the financial support of the Ph.D. scholarship of AG. AG thanks the DAAD-Germany Academic Exchange Fellowship that supports his international research visit to Computational Biophysics Laboratory, German Research School of Simulation Sciences, Germany and Professor Paolo Carloni for hospitality. Supercomputing facilities at NCI, VPAC, Swinburne University's Green Machine, Jugene (IBM Blue Gene/P) and Juropa (Intel Xeon 5570) computers in FZ Jülich must be acknowledged.

**Supporting information available:** S1: Geometry optimized structures of $Cu^{2+}$ ion in micro-solvated environment (n=1-6) using B3LY/G09 method; S2: Comparison of the average Cu-O and O-O distances in this work with experimental and other works; S3: RMSD plots of the $_l$(n=4) systems from CPMD simulations.

## References

[1] P. Sahu, S.L. Lee, Effect of microsolvation on zwitterionic glycine: an ab initio and density functional theory study. J. Mol. Model. 14 (2008) 385-392.
[2] R. Spezia, G. Tournois, J. Tortajada, T. Cartailler, M.P. Gaigeot, Toward a DFT-based molecular dynamics description of Co(ii) binding in sulfur-rich peptides. PCCP 8 (2006) 2040-2050.
[3] J. Wang, M.A. El-Sayed, The Effect of Metal Cation Binding on the Protein, Lipid and Retinal Isomeric Ratio in Regenerated Bacteriorhodopsin of Purple Membrane¶. Photochem. Photobiol. 73 (2001) 564-571.
[4] G.F. Joyce, DIRECTED EVOLUTION OF NUCLEIC ACID ENZYMES. Annu. Rev. Biochem. 73 (2004) 791-836.
[5] A.s. Binolfi, E.E. Rodriguez, D. Valensin, N. D'Amelio, E. Ippoliti, G. Obal, et al., Bioinorganic Chemistry of Parkinson's Disease: Structural Determinants for the Copper-Mediated Amyloid Formation of Alpha-Synuclein. Inorg. Chem. 49 (2010) 10668-10679.
[6] M.K. Shukla, S.K. Mishra, A. Kumar, P.C. Mishra, An ab initio study of excited states of guanine in the gas phase and aqueous media: Electronic transitions and mechanism of spectral oscillations. J. Comput. Chem. 21 (2000) 826-846.
[7] M.K. Shukla, P.C. Mishra, An ab initio study of electronic spectra and excited-state properties of 7-azaindole in vapour phase and aqueous solution. Chem. Phys. 230 (1998) 187-200.
[8] P. Rodziewicz, N.L. Doltsinis, Ab Initio Molecular Dynamics Free-Energy Study of Microhydration Effects on the Neutral–Zwitterion Equilibrium of Phenylalanine. ChemPhysChem 8 (2007) 1959-1968.
[9] J. Larrucea, E. Rezabal, T. Marino, N. Russo, J.M. Ugalde, Ab Initio Study of Microsolvated Al3+–Aromatic Amino Acid Complexes. J. Phys. Chem. B 114 (2010) 9017-9022.
[10] R. Otto, J. Brox, S. Trippel, M. Stei, T. Best, R. Wester, Single solvent molecules can affect the dynamics of substitution reactions. Nat Chem 4 (2012) 534-538.
[11] J. Bertrán, L. Rodríguez-Santiago, M. Sodupe, The Different Nature of Bonding in Cu+-Glycine and Cu2+-Glycine. J. Phys. Chem. B 103 (1999) 2310-2317.

[12] P. Hu, M.L. Gross, Gas-phase interactions of transition-metal ions and di- and tripeptides: a comparison with alkaline-earth-metal-ion interactions. J. Am. Chem. Soc. 115 (1993) 8821-8828.
[13] A. Reiter, J. Adams, H. Zhao, Intrinsic (Gas-Phase) Binding of Co2+ and Ni2+ by Peptides: A Direct Reflection of Aqueous-Phase Chemistry. J. Am. Chem. Soc. 116 (1994) 7827-7838.
[14] S. Ma, P. Wong, S.S. Yang, R.G. Cooks, Gas-Phase Molecular, Molecular Pair, and Molecular Triplet Fe+ Affinities of Pyridines. J. Am. Chem. Soc. 118 (1996) 6010-6019.
[15] B.A. Cerda, C. Wesdemiotis, The Relative Copper(I) Ion Affinities of Amino Acids in the Gas Phase. J. Am. Chem. Soc. 117 (1995) 9734-9739.
[16] A. Pasquarello, I. Petri, P.S. Salmon, O. Parisel, R. Car, É. Tóth, et al., First Solvation Shell of the Cu(II) Aqua Ion: Evidence for Fivefold Coordination. Science 291 (2001) 856-859.
[17] M. Remko, D. Fitz, R. Broer, B. Rode, Effect of metal Ions (Ni<sup>2+</sup>, Cu<sup>2+</sup> and Zn<sup>2+</sup>) and water coordination on the structure of L-phenylalanine, L-tyrosine, L-tryptophan and their zwitterionic forms. J. Mol. Model. (2011) 1-12.
[18] M.R.A. Blomberg, P.E.M. Siegbahn, S. Styring, G.T. Babcock, B. Åkermark, P. Korall, A Quantum Chemical Study of Hydrogen Abstraction from Manganese-Coordinated Water by a Tyrosyl Radical: A Model for Water Oxidation in Photosystem II. J. Am. Chem. Soc. 119 (1997) 8285-8292.
[19] A. Luna, B. Amekraz, J.P. Morizur, J. Tortajada, O. Mó, M. Yáñez, Reactions between Guanidine and Cu+ in the Gas Phase. An Experimental and Theoretical Study. J. Phys. Chem. A 101 (1997) 5931-5941.
[20] J.V. Burda, M. Pavelka, M. Simánek, Theoretical model of copper Cu(I)/Cu(II) hydration. DFT and ab initio quantum chemical study. Journal of Molecular Structure: THEOCHEM 683 (2004) 183-193.
[21] S. Amira, D. Spangberg, K. Hermansson, Distorted five-fold coordination of Cu2+(aq) from a Car-Parrinello molecular dynamics simulation. PCCP 7 (2005) 2874-2880.
[22] A. Rimola, M. Sodupe, J. Tortajada, L. Rodríguez-Santiago, Gas phase reactivity of Cu+-aromatic amino acids: An experimental and theoretical study. Int. J. Mass spectrom. 257 (2006) 60-69.
[23] M. Remko, B.M. Rode, Effect of Metal Ions (Li+, Na+, K+, Mg2+, Ca2+, Ni2+, Cu2+, and Zn2+) and Water Coordination on the Structure of Glycine and Zwitterionic Glycine. J. Phys. Chem. A 110 (2006) 1960-1967.
[24] A. Rimola, L. Rodríguez-Santiago, M. Sodupe, Cation−π Interactions and Oxidative Effects on Cu+ and Cu2+ Binding to Phe, Tyr, Trp, and His Amino Acids in the Gas Phase. Insights from First-Principles Calculations. J. Phys. Chem. B 110 (2006) 24189-24199.
[25] P.E.M. Siegbahn, Modeling aspects of mechanisms for reactions catalyzed by metalloenzymes. J. Comput. Chem. 22 (2001) 1634-1645.
[26] K.J. de Almeida, N.A. Murugan, Z. Rinkevicius, H.W. Hugosson, O. Vahtras, H. Agren, et al., Conformations, structural transitions and visible near-infrared absorption spectra of four-, five- and six-coordinated Cu(ii) aqua complexes. PCCP 11 (2009) 508-519.
[27] I. Persson, P. Persson, M. Sandstrom, A.S. Ullstrom, Structure of Jahn-Teller distorted solvated copper(ii) ions in solution, and in solids with apparently regular octahedral coordination geometry. J. Chem. Soc., Dalton Trans. (2002) 1256-1265.
[28] M. Benfatto, P. D'Angelo, S. Della Longa, N.V. Pavel, Evidence of distorted fivefold coordination of the Cu^{2+} aqua ion from an x-ray-absorption spectroscopy quantitative analysis. Physical Review B 65 (2002) 174205.
[29] P. Frank, M. Benfatto, R.K. Szilagyi, P. D'Angelo, S.D. Longa, K.O. Hodgson, The Solution Structure of [Cu(aq)]2+ and Its Implications for Rack-Induced Bonding in Blue Copper Protein Active Sites. Inorg. Chem. 44 (2005) 1922-1933.
[30] J. Chaboy, A. Munoz-Paez, P.J. Merkling, E.S. Marcos, The hydration of Cu[sup 2+]: Can the Jahn-Teller effect be detected in liquid solution? J. Chem. Phys. 124 (2006) 064509-064509.

[31] M. Nomura, T. Yamaguchi, Concentration dependence of EXAFS and XANES of copper(II) perchlorate aqueous solution: comparison of solute structure in liquid and glassy states. The Journal of Physical Chemistry 92 (1988) 6157-6160.
[32] D.H. Powell, L. Helm, A.E. Merbach, [sup 17]O nuclear magnetic resonance in aqueous solutions of Cu[sup 2 + ] : The combined effect of Jahn--Teller inversion and solvent exchange on relaxation rates. J. Chem. Phys. 95 (1991) 9258-9265.
[33] J. Garcia, M. Benfatto, C.R. Natoli, A. Bianconi, A. Fontaine, H. Tolentino, The quantitative Jahn-teller distortion of the Cu2+ site in aqueous solution by xanes spectroscopy. Chem. Phys. 132 (1989) 295-302.
[34] B. Beagley, A. Eriksson, J. Lindgren, I. Persson, L.G.M. Pettersson, M. Sandstrom, et al., A computational and experimental study on the Jahn-Teller effect in the hydrated copper (II) ion. Comparisons with hydrated nickel (II) ions in aqueous solution and solid Tutton's salts. J. Phys.: Condens. Matter 1 (1989) 2395.
[35] P.S. Salmon, G.W. Neilson, J.E. Enderby, The structure of Cu 2+ aqueous solutions. Journal of Physics C: Solid State Physics 21 (1988) 1335-1349.
[36] K. Ozutsumi, T. Kawashima, Exafs and spectrophotometric studies on the structure of pyridine complexes with copper(II) and copper(I) ions in aqueous solution. Polyhedron 11 (1992) 169-175.
[37] M. Sano, T. Maruo, Y. Masuda, H. Yamatera, Structural study of copper(II) sulfate solution in highly concentrated aqueous ammonia by x-ray absorption spectra. Inorg. Chem. 23 (1984) 4466-4469.
[38] M. Valli, S. Matsuo, H. Wakita, T. Yamaguchi, M. Nomura, Solvation of Copper(II) Ions in Liquid Ammonia. Inorg. Chem. 35 (1996) 5642-5645.
[39] M. Valli, S. Matsuo, H. Wakita, T. Yamaguchi, M. Nomura, ChemInform Abstract: Solvation of Copper(II) Ions in Liquid Ammonia. ChemInform Abstract 28 (1997).
[40] J. Emsley, M. Arif, P.A. Bates, M.B. Hursthouse, Diaquabis(1,3-diaminopropane)copper(II) difluoride: X-ray structure reveals short hydrogen bonds between ligand waters and lattice fluorides. Inorg. Chim. Acta 154 (1988) 17-20.
[41] J. Emsley, M. Arif, P.A. Bates, M.B. Hursthouse, Hydrogen bonding between free fluoride ions and water molecules: two X-ray structures. J. Mol. Struct. 220 (1990) 1-12.
[42] A.A.G. Tomlinson, B.J. Hathaway, The electronic properties and stereochemistry of the copper(II) ion. Part III. Some penta-ammine complexes. Journal of the Chemical Society A: Inorganic, Physical, Theoretical (1968) 1905-1909.
[43] M. Duggan, N. Ray, B. Hathaway, G. Tomlinson, P. Brint, K. Pelin, Crystal structure and electronic properties of ammine[tris(2-aminoethyl)amine]copper(II) diperchlorate and potassium penta-amminecopper(II) tris(hexafluorophosphate). J. Chem. Soc., Dalton Trans. (1980) 1342-1348.
[44] H. Elliott, B.J. Hathaway, The Hexaammine Complexes of the Copper(II) Ion. Inorg. Chem. 5 (1966) 885-889.
[45] T.M. Distler, P.A. Vaughan, Crystal structures of the hexaamminecopper(II) halides. Inorg. Chem. 6 (1967) 126-129.
[46] L.r. Rulíšek, J. Vondrášek, Coordination geometries of selected transition metal ions (Co2+, Ni2+, Cu2+, Zn2+, Cd2+, and Hg2+) in metalloproteins. J. Inorg. Biochem. 71 (1998) 115-127.
[47] J. Gonzalez, E. Florez, J. Romero, A. Reyes, A. Restrepo, Microsolvation of Mg2+, Ca2+: strong influence of formal charges in hydrogen bond networks. J. Mol. Model. (2013) 1-15.
[48] C. Ibarguen, M. Manrique-Moreno, C.Z. Hadad, J. David, A. Restrepo, Microsolvation of dimethylphosphate: a molecular model for the interaction of cell membranes with water. PCCP 15 (2013) 3203-3211.
[49] Z. Huang, W. Yu, Z. Lin, Exploration of the full conformational landscapes of gaseous aromatic amino acid phenylalanine: An ab initio study. Journal of Molecular Structure: THEOCHEM 758 (2006) 195-202.

[50] U. Purushotham, D. Vijay, G. Narahari Sastry, A computational investigation and the conformational analysis of dimers, anions, cations, and zwitterions of L-phenylalanine. J. Comput. Chem. 33 (2012) 44-59.
[51] L.C. Snoek, R.T. Kroemer, M.R. Hockridge, J.P. Simons, Conformational landscapes of aromatic amino acids in the gas phase: Infrared and ultraviolet ion dip spectroscopy of tryptophan. PCCP 3 (2001) 1819-1826.
[52] G. Schaftenaar, J.H. Noordik, Molden: a pre- and post-processing program for molecular and electronic structures*. J. Comput.-Aided Mol. Des. 14 (2000) 123-134.
[53] W. Humphrey, A. Dalke, K. Schulten, VMD: Visual molecular dynamics. J. Mol. Graphics 14 (1996) 33-38.
[54] M.J. Frisch, G.W. Trucks, H.B. Schlegel, G.E. Scuseria, M.A. Robb, J.R. Cheeseman, et al., Gaussian 09, Revision A.02. Wallingford CT, (2009).
[55] P.J. Hay, Gaussian basis sets for molecular calculations. The representation of 3d orbitals in transition-metal atoms. J. Chem. Phys. 66 (1977) 4377-4384.
[56] A.J.H. Wachters, Gaussian Basis Set for Molecular Wavefunctions Containing Third-Row Atoms. J. Chem. Phys. 52 (1970) 1033-1036.
[57] A. Ganesan, F. Wang, Intramolecular interactions of L-phenylalanine revealed by inner shell chemical shift. J. Chem. Phys. 131 (2009) 044321-044329.
[58] A. Ganesan, F. Wang, C. Falzon, Intramolecular interactions of L-phenylalanine: Valence ionization spectra and orbital momentum distributions of its fragment molecules. J. Comput. Chem. 32 (2011) 525-535.
[59] C.T. Falzon, F. Wang, W. Pang, Orbital Signatures of Methyl in l-Alanine. J. Phys. Chem. B 110 (2006) 9713-9719.
[60] M.R.A. Blomberg, P.E.M. Siegbahn, M. Svensson, Comparisons of results from parametrized configuration interaction (PCI-80) and from hybrid density functional theory with experiments for first row transition metal compounds. J. Chem. Phys. 104 (1996) 9546-9554.
[61] M.C. Holthausen, M. Mohr, W. Koch, The performance of density functional/Hartree-Fock hybrid methods: the bonding in cationic first-row transition metal methylene complexes. Chem. Phys. Lett. 240 (1995) 245-252.
[62] C. Adamo, F. Lelj, A hybrid density functional study of the first-row transition-metal monocarbonyls. J. Chem. Phys. 103 (1995) 10605-10613.
[63] R. Car, M. Parrinello, Unified Approach for Molecular Dynamics and Density-Functional Theory. Phys. Rev. Lett. 55 (1985) 2471.
[64] v. CPMD, C. (revision a11); Copyright IBM Corp, 1990−2008; Copyright MPI fr Festkrperforschung Stuttgart, 1997−2001; http://www.cpmd.org/.
[65] A.D. Becke, Density-functional exchange-energy approximation with correct asymptotic behavior. Physical Review A 38 (1988) 3098-3100.
[66] C. Lee, W. Yang, R.G. Parr, Development of the Colle-Salvetti correlation-energy formula into a functional of the electron density. Physical Review B 37 (1988) 785.
[67] N. Troullier, J. Martins, eacute, Luriaas, Efficient pseudopotentials for plane-wave calculations. Physical Review B 43 (1991) 1993.
[68] S.G. Louie, S. Froyen, M.L. Cohen, Nonlinear ionic pseudopotentials in spin-density-functional calculations. Physical Review B 26 (1982) 1738.
[69] S. Nose, A unified formulation of the constant temperature molecular dynamics methods. J. Chem. Phys. 81 (1984) 511-519.
[70] W.G. Hoover, Canonical dynamics: Equilibrium phase-space distributions. Physical Review A 31 (1985) 1695.
[71] Z. Huang, W. Yu, Z. Lin, Exploration of the full conformational landscapes of gaseous aromatic amino acid phenylalanine: An ab initio study. Journal of Molecular Structure: THEOCHEM 758 (2006) 195-202.

[72] A. Bérces, T. Nukada, P. Margl, T. Ziegler, Solvation of Cu2+ in Water and Ammonia. Insight from Static and Dynamical Density Functional Theory. J. Phys. Chem. A 103 (1999) 9693-9701.
[73] M. Magini, Coordination of copper(II). Evidence of the Jahn-Teller effect in aqueous perchlorate solutions. Inorg. Chem. 21 (1982) 1535-1538.
[74] A. Musinu, G. Paschina, G. Piccaluga, M. Magini, Coordination of copper(II) in aqueous copper sulfate solution. Inorg. Chem. 22 (1983) 1184-1187.
[75] M. Remko, D. Fitz, B. Rode, Effect of metal ions (Li+, Na+, K+, Mg2+, Ca2+, Ni2+, Cu2+ and Zn2+) and water coordination on the structure and properties of l-histidine and zwitterionic l-histidine. Amino Acids 39 (2010) 1309-1319.
[76] L. Julen, Car–Parrinello molecular dynamics study of the coordination on Al 3+ (aq). Phys. Scr. 84 (2011) 045305.
[77] H. Wincel, Hydration Energies of Sodiated Amino Acids from Gas-Phase Equilibria Determinations. J. Phys. Chem. A 111 (2007) 5784-5791.
[78] M.A. Addicoat, G.F. Metha, T.W. Kee, Density functional theory investigation of Cu(I)- and Cu(II)-curcumin complexes. J. Comput. Chem. 32 (2011) 429-438.

**Table 1:** Selected geometrical parameters of the $[Phe\text{-}Cu]^{2+}$ complexes optimized using the B3LYP/G09 and BLYP/CPMD methods.

| Parameters[$] | $[Phe1\text{-}Cu]^{2+}$ (NT) | | | $[Phe2\text{-}Cu]^{2+}$ (NT) | | $[Phe3\text{-}Cu]^{2+}$ (NT) | | $[Phe4\text{-}Cu]^{2+}$ (ZW) | | | $[Phe5\text{-}Cu]^{2+}$ (ZW) | | $[Phe5a\text{-}Cu]^{2+}$ (ZW) | |
|---|---|---|---|---|---|---|---|---|---|---|---|---|---|---|
| | B3LYP[a] | BLYP[b] | Other Work[#] | B3LYP[a] | BLYP[b] | B3LYP[a] | BLYP[b] | B3LYP[a] | BLYP[b] | Other Work[#] | B3LYP[a] | BLYP[b] | B3LYP[a] | BLYP[b] |
| $C_{(1)}$-$C_{(2)}$/Å | 1.55 | 1.55 | 1.55 | 1.54 | 1.55 | 1.57 | 1.59 | 1.57 | 1.57 | 1.57 | 1.57 | 1.59 | 1.52 | 1.53 |
| $C_{(2)}$-$O_{(3)}$/ Å | 1.23 | 1.25 | 1.22 | 1.24 | 1.25 | 1.28 | 1.30 | 1.24 | 1.24 | 1.23 | 1.30 | 1.23 | 1.29 | 1.30 |
| $C_{(2)}$-$O_{(4)}$/ Å | 1.31 | 1.32 | 1.31 | 1.30 | 1.32 | 1.25 | 1.26 | 1.28 | 1.30 | 1.27 | 1.22 | 1.31 | 1.25 | 1.26 |
| $C_{(1)}$-N/ Å | 1.49 | 1.53 | 1.49 | 1.50 | 1.51 | 1.44 | 1.44 | 1.52 | 1.55 | 1.52 | 1.52 | 1.54 | 1.52 | 1.54 |
| $O_{(4)}$-Cu/ Å | 2.05 | 2.03 | 2.08 | 2.06 | 2.12 | 1.87 | 1.93 | 1.91 | 1.92 | 1.92 | 1.82 | 1.88 | 2.11 | 2.24 |
| $O_{(3)}$-Cu/ Å | – | – | – | – | – | – | – | 2.98 | 2.90 | 3.02 | – | – | 2.01 | 2.06 |
| N-Cu/ Å | 2.08 | 2.07 | 2.11 | 2.03 | 2.09 | – | – | – | – | – | – | – | – | – |
| ∠$C_{(1)}$-$C_{(2)}$-$O_{(3)}$/° | 123.20 | 123.40 | 122.90 | 119.02 | 119.50 | 116.30 | 116.80 | 114.60 | 119.10 | 116.50 | 117.40 | 117.60 | 117.50 | 117.10 |
| ∠$C_{(2)}$-$C_{(1)}$-N/° | 109.60 | 109.30 | 109.00 | 104.70 | 105.10 | 107.00 | 107.90 | 103.70 | 104.30 | 103.80 | 103.60 | 102.80 | 109.60 | 109.50 |
| ∠$C_{(1)}$-N-Cu/° | 108.50 | 107.60 | 109.30 | 99.60 | 99.10 | – | – | – | – | – | – | – | – | – |
| ∠$C_{(2)}$-$O_{(4)}$-Cu/° | 112.40 | 111.80 | 113.40 | 108.30 | 108.50 | 131.30 | 128.60 | 114.90 | 112.70 | 117.30 | 132.90 | 128.70 | 81.90 | 80.50 |
| ∠$C_{(2)}$-$O_{(3)}$-Cu/° | – | – | – | – | – | – | – | 65.80 | 68.90 | 65.30 | – | – | 85.10 | 86.90 |
| ∠$O_{(3)}$-Cu-N/° | 83.60 | 85.70 | 81.60 | 80.20 | 79.20 | – | – | – | – | – | – | – | – | – |
| ∠$O_{(3)}$-Cu-$O_{(4)}$/° | – | – | – | – | – | – | – | 49.50 | 51.60 | 48.20 | – | – | 64.70 | 62.20 |
| ∠$O_{(3)}$-$C_{(2)}$-$C_{(1)}$-N/° | 12.00 | 11.70 | 15.20 | -30.70 | -32.70 | 2.40 | -1.90 | 7.70 | 7.20 | 7.90 | -7.30 | -10.70 | -45.10 | -41.60 |
| ∠$C_{(2)}$-$C_{(1)}$-N-Cu/° | -16.80 | -15.10 | -20.30 | 50.60 | 51.50 | – | – | – | – | – | – | – | – | – |
| ∠$C_{(1)}$-$C_{(2)}$-$O_{(4)}$-Cu/° | 0.20 | -1.20 | -0.90 | -7.60 | -6.30 | 5.60 | 8.20 | -177.00 | -177.10 | -177.00 | 22.40 | 22.80 | 134.50 | 131.80 |
| ∠$C_{(1)}$-$C_{(2)}$-$O_{(3)}$-Cu/° | – | – | – | – | – | – | – | 177.60 | 177.60 | 176.60 | – | – | -133.70 | -129.80 |
| $R_6$[^] /Å | 8.48 | 8.48 | | 8.49 | 8.53 | 8.51 | 8.54 | 8.48 | 8.48 | | 8.50 | 8.54 | 8.48 | 8.52 |

[$]Here $C_{(1)}$ represents the $C_{(\alpha)}$; $C_{(2)}$ represents carbonyl carbon (COO); $O_{(3)}$ and $O_{(4)}$ denote the carbonyl oxygen atoms. Refer to Fig.1 for atom numbering.

[a]B3LYP/6-311++G(d,p) model using G09;

[b]BLYP with MT pseudopotentials using CPMD;

[#]Ref 17 (B3LYP/6-311+G(d,p) [17].

^$R_6$ is the perimeter of the Phenyl ring (The values in the table are to be compared with the $R_6$ of Phe (8.37) [58]).

**Table 2:** Selected distances of $Cu^{2+}$ in the lowest energy micro-hydrated complexes (in Å).

| Parameters | $_{l}$[Phe1-Cu]$^{2+}$ (NT) | $_{l}$(n=1)$^{\$}$ (NT) | $_{l}$(n=2)$^{\$}$ (ZW) | $_{l}$(n=3)$^{\$}$ (ZW) | $_{l}$(n=4)$^{\$}$ (ZW) |
|---|---|---|---|---|---|
| | B3LYP[a] (BLYP[b]) | B3LYP[a] (BLYP[b]) | B3LYP[a] (BLYP[b]) | B3LYP[a] (BLYP[b]) | B3LYP[a] (BLYP[b]) |
| **Cu-O$_{(3)}$** | | | 2.02 (2.18) | 3.28 (3.33) | 3.23 (3.30) |
| **Cu-O$_{(4)}$** | 2.05 (2.03) | 2.09 (2.12) | 1.97 (2.08) | 1.91 (2.02) | 1.95 (2.04) |
| **Cu-N** | 2.08 (2.07) | 2.01 (2.03) | | | |
| **Cu-O$_{(wat1)}$** | | 1.94 (1.98) | 1.98 (2.06) | 2.02 (2.12) | 2.06 (2.15) |
| **Cu-O$_{(wat2)}$** | | | 1.98 (2.09) | 1.99 (2.09) | 1.96 (2.07) |
| **Cu-O$_{(wat3)}$** | | | | 1.95 (2.04) | 2.01 (2.13) |
| **Cu-O$_{(wat4)}$** | | | | | 2.31 (2.39) |

$^{\$}$ $_{l}$(n=1) denotes $_{l}$[Phe1-Cu(n=1)]$^{2+}$; $_{l}$(n=2) denotes $_{l}$[Phe4-Cu(n=2)]$^{2+}$;
$_{l}$(n=3) denotes $_{l}$[Phe5-Cu(n=3)]$^{2+}$; $_{l}$(n=4) denotes $_{l}$[Phe5a-Cu(n=4)]$^{2+}$ .
[a]B3LYP/6-31++G** model using G09;
[b]BLYP with MT pseudopotentials using CPMD.

**Table 3:** Selected geometrical parameters of the initial and final (shaded in grey color) snapshots of the micro-hydrated complexes in the CPMD simulations.

| Geometrical Parameters | $_l$[Phe1-Cu(n=1)]$^{2+}$ (NT) | | $_l$[Phe4-Cu(n=2)]$^{2+}$ (ZW) | | $_l$[Phe5-Cu(n=3)]$^{2+}$ (ZW) | | $_l$[Phe5a-Cu(n=4)]$^{2+}$ (ZW) | |
|---|---|---|---|---|---|---|---|---|
| | **Initial** | **Final** | **Initial** | **Final** | **Initial** | **Final** | **Initial** | **Final** |
| d[Cu-$O_{(3)}$]/Å | | | 2.18 | 3.19 | 3.33 | 3.53 | 3.30 | 3.57 |
| d[Cu-$O_{(4)}$]/Å | 2.12 | 2.03 | 2.08 | 1.94 | 2.02 | 2.45 | 2.04 | 1.98 |
| d[Cu-N)]/Å | 2.03 | 2.14 | | | | | | |
| d[Cu-$O_{(wat1)}$]/Å | 1.98 | 2.09 | 2.06 | 2.07 | 2.12 | 2.25 | 2.15 | 2.37 |
| d[Cu-$O_{(wat2)}$]/Å | | | 2.09 | 2.00 | 2.09 | 2.11 | 2.07 | 2.08 |
| d[Cu-$O_{(wat3)}$]/Å | | | | | 2.04 | 1.98 | 2.13 | 2.16 |
| d[Cu-$O_{(wat4)}$]/Å | | | | | | | 2.39 | 4.30 |
| **Hydrogen bonds** | | | | | | | | |
| (N)H···$O_{(3)}$*/Å | 3.31 | 3.28 | <u>2.19</u> | <u>2.03</u> | <u>1.91</u> | <u>1.61</u> | <u>1.90</u>^ | <u>1.98</u>^ |
| $H_{(wat3)}$···$O_{(3)}$*/Å | 4.29 | 4.84 | 3.36$ | <u>2.79</u>$ | <u>1.72</u> | <u>1.74</u> | <u>1.67</u> | <u>2.73</u> |
| $H_{(wat1)}$···$O_{(wat2)}$*/Å | - | - | - | - | - | - | 3.32 | <u>2.47</u> |
| $H_{(wat2)}$···$O_{(wat4)}$*/Å | - | - | - | - | - | - | 3.67 | <u>1.54</u> |
| $Cu^{2+}$ coordination | 3 | 3 | 4 | 3 | 4 | 4 | 5 | 4 |

*Hydrogen bonds within the complexes. Hydrogen bonds within 2.80 Å are underlined.

$ $H_{(wat2)}$···$O_{(3)}$ bond in $_l$[Phe4-Cu(n=2)]$^{2+}$complex.

^The H-bond in the initial structure of $_l$[Phe5a-Cu(n=4)]$^{2+}$complex is between $O_{(3)}$ and $H_x$ in $NH_3$, whereas in the final structure, the H-bond is between $O_{(3)}$ and $H_Y$ in $NH_3$.

**Figure Captions:**

**Fig. 1.** The optimized structures of phenylalanine (Phe), $[Phe\text{-}Cu]^{2+}$ and micro-hydrated $[Phe\text{-}Cu(n=0\text{-}4)]^{2+}$ structures, along with their relative energies in kcal·mol$^{-1}$ obtained from B3LYP/6-311++G(d,p) calculations. Here n represents the number of water molecules in the system. The lowest energy structures (***l***) are indicated in boxes.

**Fig. 2**. Last snapshot of micro-solvated $[Phe\text{-}Cu]^{2+}$ structures from the 12 ps CPMD simulation. after the CPMD simulations. The inter- and intra-molecular hydrogen bonds are also indicated (Cu coordinated bonds are shown as solid lines and the dotted lines represent the H-Bonds).

**Fig. 3**. Trajectories of the (a) Cu-O distances and (b) inter-molecular H-O distances and (c) intra-molecular $NH\text{-}O_{(3)}$ distances of the $_{l}[Phe5a\text{-}Cu(n=4)]^{2+}$ complex.

**Fig. 4**. Radial distribution function spectra for the lowest energy micro-hydrated $[Phe\text{-}Cu]^{2+}$ complexes calculated from the CPMD trajectories.

**Fig. 1**

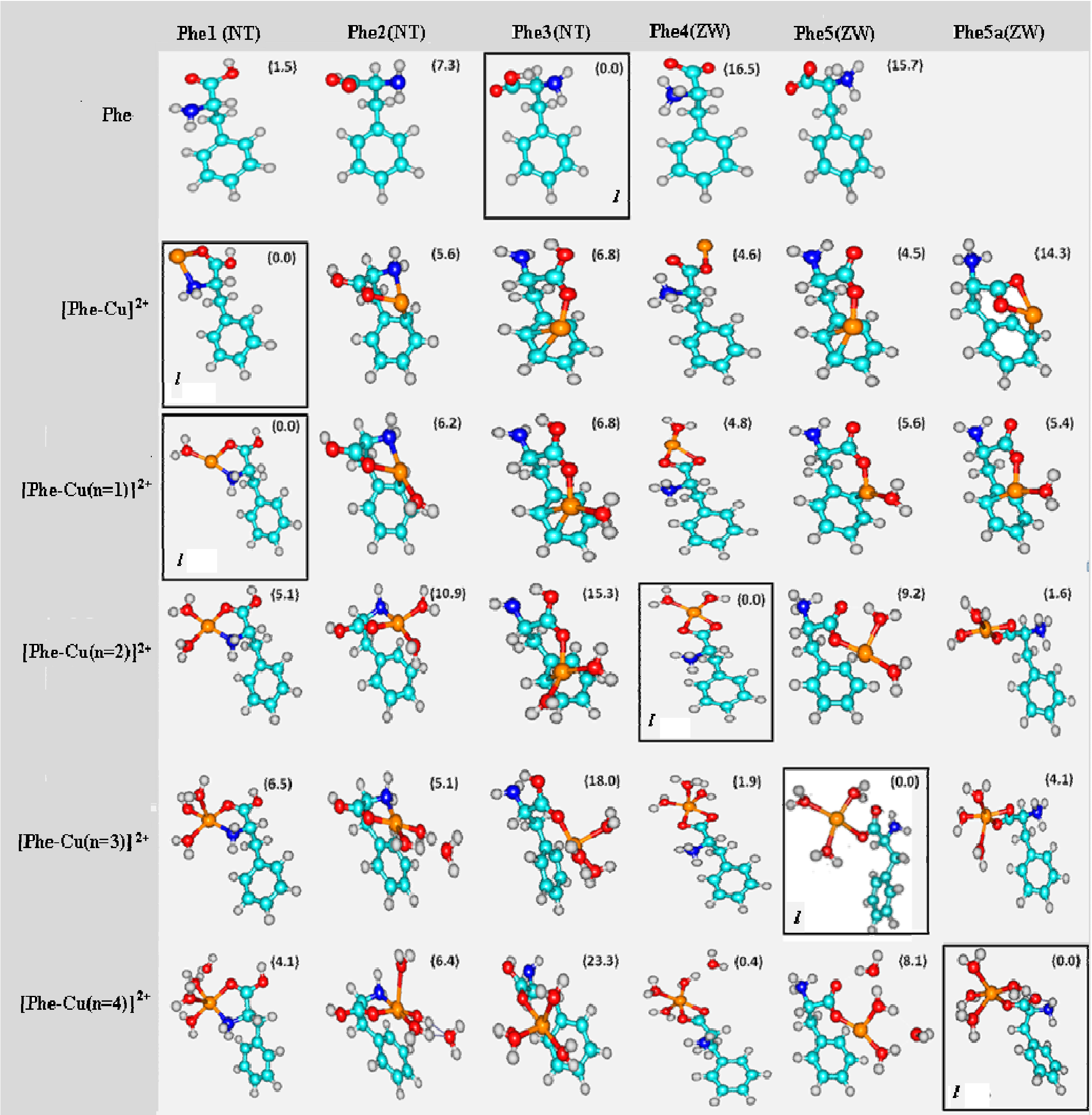

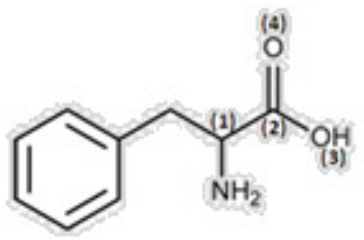

NT form

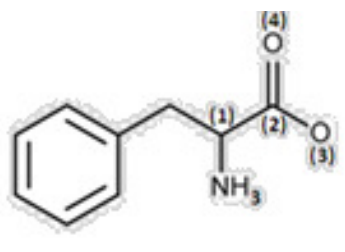

ZW form

**Fig.**

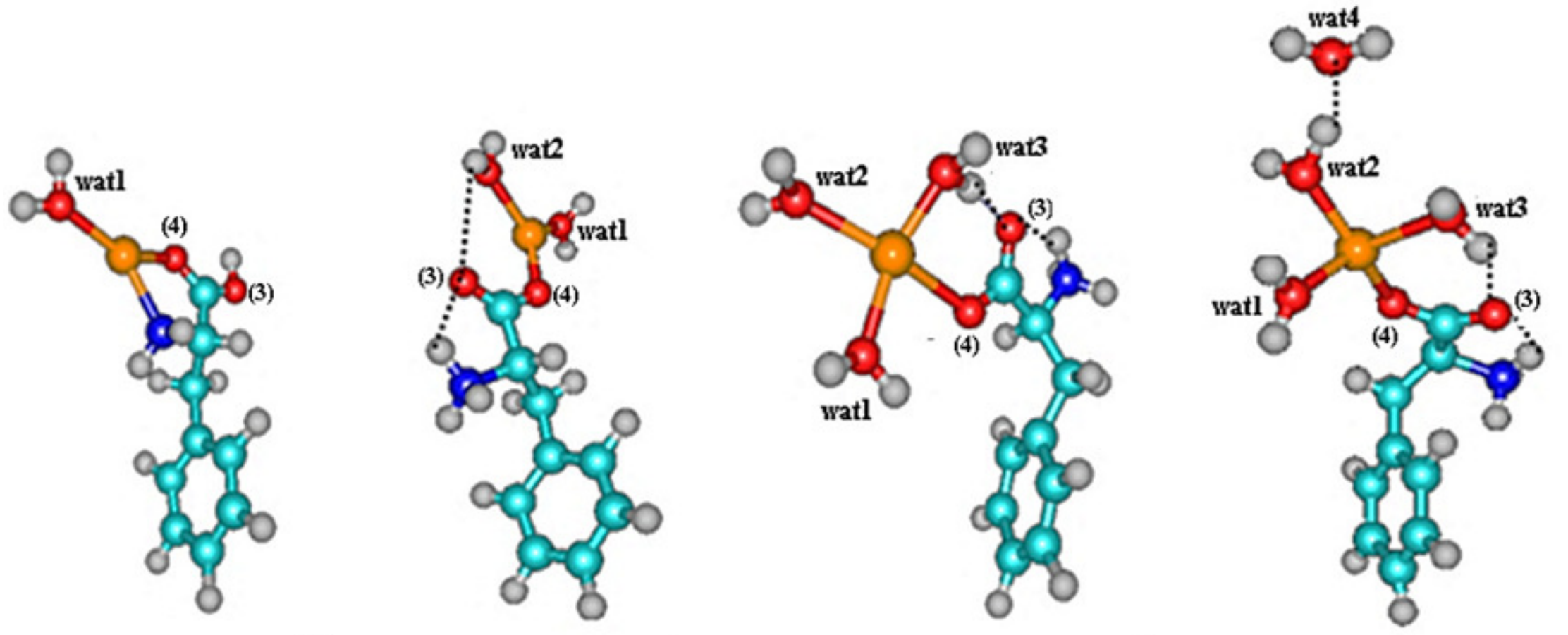

**2** $_l$[Phe1-Cu(n=1)]$^{2+}$ (a) $_l$[Phe4-Cu(n=2)]$^{2+}$ (b) $_l$[Phe5-Cu(n=3)]$^{2+}$ (c) $_l$[Phe5a-Cu(n=4)]$^{2+}$ (d)

**Fig. 3**

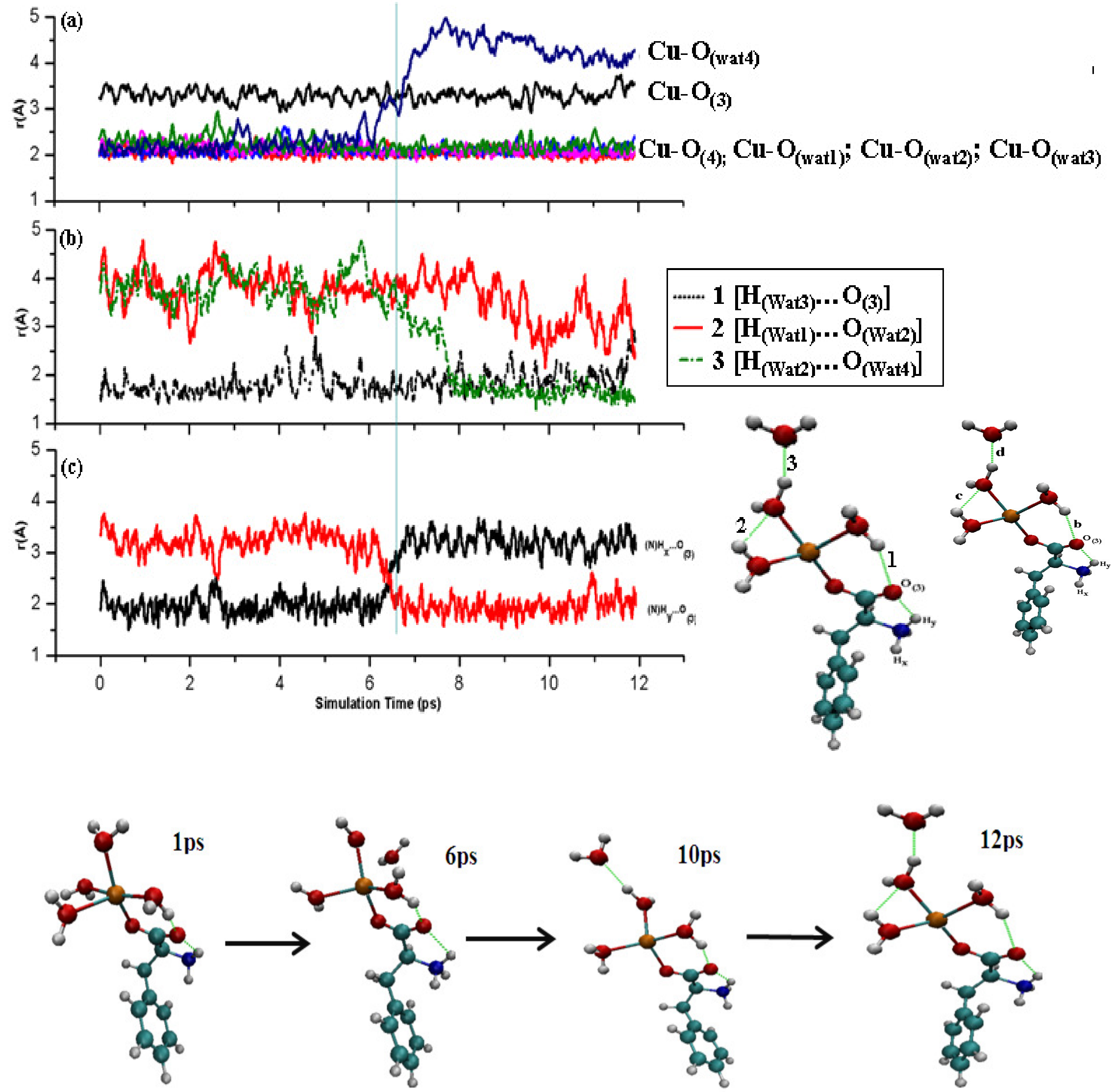

**Fig. 4**

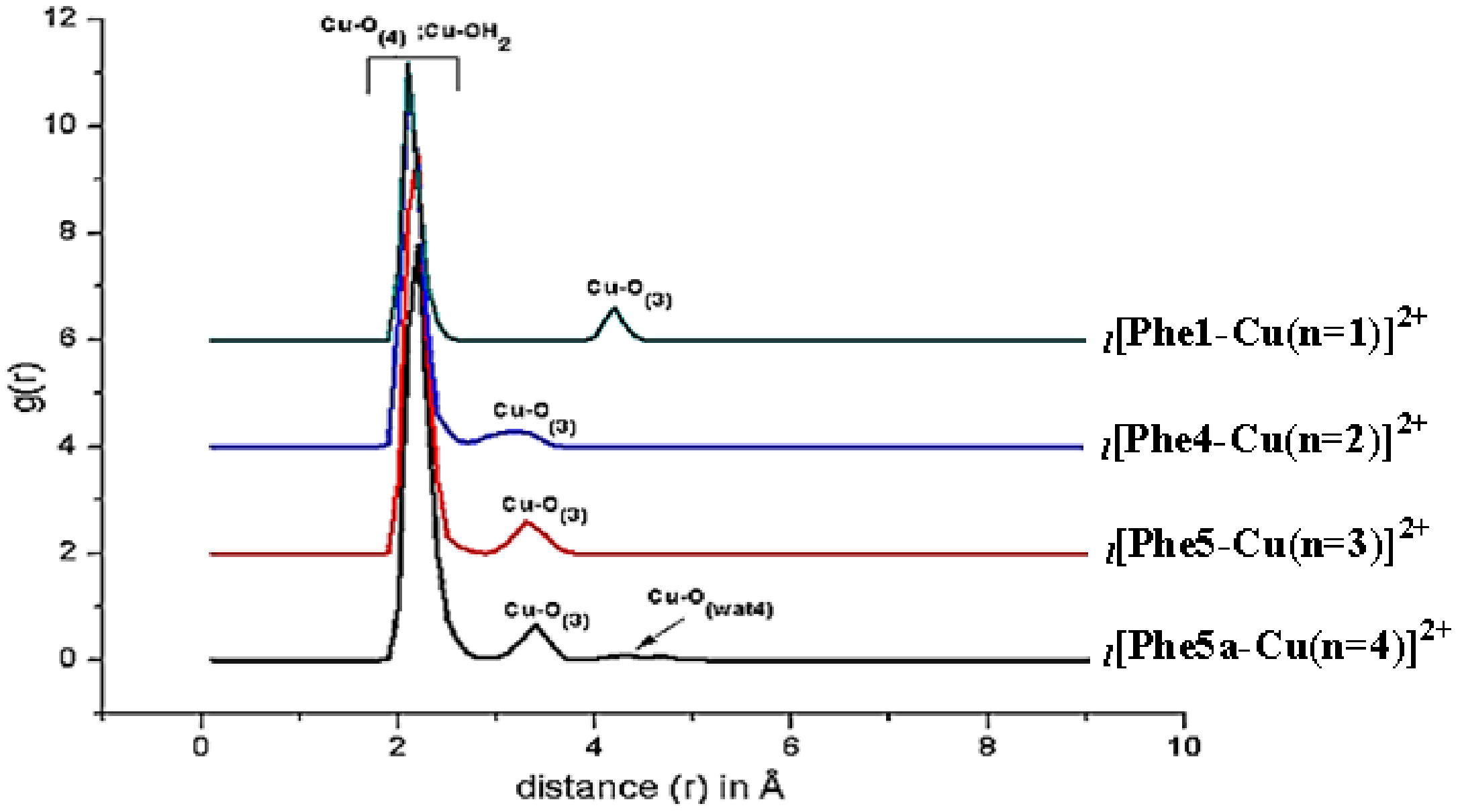